\documentclass{ws-p10x7}
\catcode`\@=11
     \def\@cite#1#2{{[#1]\if@tempswa\typeout
     	{WSPC warning: optional citation argument 
     	ignored: `#2'} \fi}}
\def\ps@headings{}
\catcode`@=12
\def\cropmarks{}
\newcommand\Invtalk{\begin{center} \it
Invited talk given at the XX International Symposium \\ on
Lepton and Photon Interactions at High Energies \\(Rome, Italy, 23-28
July 2001).\end{center}\vskip 0.3cm}

\newcommand\figsize{0.45\textwidth}
\newcommand\newblock{}

\newcommand     \ba             {\begin{eqnarray}}
\newcommand     \be             {\begin{equation}}
\newcommand     \Ca             {{C_{\rm A}}}
\newcommand     \Cf             {{C_{\rm F}}}

\newcommand     \ea             {\end{eqnarray}}
\newcommand     \ee             {\end{equation}}

\newcommand     \epem           {\ifmmode{e^+e^-}\else{$e^+e^-$}\fi}

\newcommand     \lambdamsb     {\ifmmode
          \Lambda_5^{\rm \scriptscriptstyle \overline{MS}} \else
         $\Lambda_5^{\rm \scriptscriptstyle \overline{MS}}$ \fi}
\newcommand     \Lambdamsb      \lambdamsb
\newcommand     \LambdaQCD     {\ifmmode
          \Lambda_{\rm \scriptscriptstyle QCD} \else
         $\Lambda_{\rm \scriptscriptstyle QCD}$ \fi}

\newcommand     \MSB            {\ifmmode {\overline{\rm MS}} \else
                                 $\overline{\rm MS}$  \fi}

\newcommand     \nf             {n_{\rm f}}

\newcommand     \ptmin     {\ifmmode p_{\scriptscriptstyle T}^{\sss min} \else
                           $p_{\scriptscriptstyle T}^{\sss min}$ \fi}

\newcommand     \Tf             {{T_{\rm F}}}

\newcommand\sss{\scriptscriptstyle\rm}

\newcommand\as{\alpha_{\sss S}}

\def \pt   {p_{\scriptscriptstyle T}}

\def \to   {\mbox{$\rightarrow$}}

\newcommand\spreadlines[1]{}
\newcommand\red{}
\newcommand\blue{}

\begin{document}
\author{Paolo Nason}
\title{QCD at high energy}
\address{INFN, sez. di Milano,\\ and Universit\`a di Milano-Bicocca}
\twocolumn[\maketitle\Invtalk\abstract{
I review few recent QCD results in $e^+e^-$, $ep$ and $p\bar{p}$
collisions.
Furthermore, I discuss recent studies in power suppressed effects,
ongoing progress in next-to-next-to-leading QCD calculations,
and some recent puzzling results in $b$ production.
}]
\section{Introduction}
Much of the present and future of high energy physics relies on our
ability to understand strong interactions in the high energy regime.
While there is widespread believe that QCD is the right description
of the strongly interacting world, a solution of the theory is not yet
available, and its applications require further assumptions. The perturbative
framework, in particular, is based upon the assumption that quantities
that can be reliably computed in perturbation theory
(i.e. infrared safe quantities)
do correctly predict corresponding measurable quantities, up to effects
which are suppressed by inverse powers of the characteristic energy scale
of the process. This assumption (often called parton-hadron
duality), implies that a description of a phenomenon in terms
of constituents correctly describes the behaviour of the hadrons
that form the final state. The properties of the QCD Lagrangian are
such that this assumption is consistent, i.e. it implies no contradictions.
Thus, for example, perturbative QCD does not allow us to compute
the cross section for the production of an isolated quark or gluon,
consistently with the fact that no isolated quark and gluons are observed
in nature. It remains, however, and unproven assumption, and only
extensive testing can give us confidence
on its validity.

QCD studies at high energy aim at giving us convincing evidence
of the viability of this approach. By now, we have accumulated a wide
body of experimental evidence in favour of it.
The three main experimental areas for QCD tests
are hadron production in $e^+e^-$ annihilation, deep inelastic scattering,
and high $\pt$ production phenomena in hadronic collisions.

LEP has proven to be an ideal experiment for testing perturbative
QCD. In some sense, the $e^+e^-$ annihilation environment is the most
appropriate for studying jet production, since the initial state
is fully known. Systematic studies of large classes of jet shape variables
have been carried out, to such an extent that one can no longer
doubt that perturbative QCD is at work in $e^+e^-$ annihilation.

The $ep$ collision environment has traditionally provided
an area of QCD studies in the framework of structure
functions and scaling violation. Besides this, the high energies
available at HERA have provided a new environment where to study
hard production phenomena.

The highest scales currently available for QCD studies are reached
at hadron colliders.
The applications of QCD to hadronic collisions go beyond the pure problem
of testing. The $W$ and $Z$ vector bosons, as well as the $t\bar{t}$
production cross sections were computed in the perturbative
QCD framework, and represent remarkable examples in which perturbative
QCD has allowed to predict the cross section for unknown particles.
Higgs and Susy searches at the Tevatron and at the LHC also rely on
our ability to compute the production cross sections.
Thus, reaching true precision in our understanding
of hard production phenomena becomes a fundamental issue if we want
to search for new physics.

In the past twenty
years, a large effort has gone into refining theoretical calculations,
as well as experimental analysis, in order to perform meaningful
tests of the theory. Most of the theoretical effort has gone
into computing next-to-leading order (NLO) corrections to the parton model
predictions of QCD.
The impressive success of LEP jet physics
would not have been possible without the inclusion of such corrections.
Further theoretical progress has been made in the 
all order resummation of enhanced contributions due to soft
gluon emission, and in the study of the high-energy (small-$x$)
regime.
Currently, some effort is being made in constructing models of power
suppressed effects that have better theoretical motivations than
Monte Carlo hadronization models.

There are several indications that knowledge of next-to-next-to-leading
corrections would further improve our understanding of QCD.
In jet physics at LEP, a large part of the theoretical error is due to
scale dependence, that is to say, to unknown higher order effects.
In hadron collider physics, there are several instances where
one finds large radiative corrections, of the order of 100\%\ of
the parton model prediction. This is the case, for example,
of bottom pair production and Higgs production. Going one step further
in the precision of QCD calculation would be invaluable help
in understanding and modeling hadron collider processes. Remarkable
theoretical progress has taken place in this past years, that suggests
that going beyond NLO will indeed be possible in the near future.

In the following sections I will review recent progress in these fields.
Completeness, in this short review, is not possible, and thus I
apologize for leaving out many important developments.
I will not discuss deep inelastic scattering
and small-$x$ physics topics, since they will be covered in
\cite{Erdmann:LP2001,Carli:LP2001}.
I will pick few examples in $e^+e^-$, $ep$ and $p\bar{p}$ physics
that illustrate the current status of QCD studies.
I will discuss recent power
corrections studies, and I will describe the current
theoretical effort that is being made for reaching NNLO results
in QCD. I will also discuss some recent puzzling results in $b$ production,
which represents an area where relevant discrepancies with QCD predictions
are observed.
\section{QCD in $e^+e^-\to\mbox{hadrons}$}
\subsection{Theoretical basis}
The theoretical basis of QCD studies at LEP mainly relies on the
calculation of ref.~\cite{Ellis:1981wv}, which allows to compute
any infrared safe $3-$jets shape variable as an expansion
of the form
$\as(\mu^2)A+\as^2(\mu^2) B(\mu^2/Q^2)+\ldots$.
Heavy quark mass effects have also been included in the $3-$jets
calculation \cite{Rodrigo:1999qg,Brandenburg:1998pu,Nason:1998nw}.

Recently, the NLO correction to four partons production have been
computed
\cite{Dixon:1997th,Nagy:1997yn,Campbell:1998nn,Weinzierl:1999yf},
allowing thus the computation of any $4-$jets shape variable
in the form $\as^2(\mu^2)C+\as^3(\mu^2)D(\mu^2/Q^2)+\ldots$.

In the limit of thin jets, Sudakov logarithm arise to all order
in the perturbative expansions.
In some cases, these logarithms can be organized and resummed
in the following form
\cite{Catani:1991kz,Catani:1991hj,Catani:1993ua}
\begin{displaymath}
R(y)= F(\as)\,e^{Lg_1(\as L)+g_2(\as L)}
\end{displaymath}
where {\red $R$} is a jet rate, {\red $y$} the jet ``thickness'',
{\red $L=\log 1/y$}. The term $g_1$, which is enhanced by a logarithmic
factor with respect to $g_2$, is the leading-log term (LL),
and the $g_2$ term is the next-to-leading-log term (NLL).

Hadronization and power corrections are believed to be
suppressed as $1/Q$, but they are still important at LEP energies.
They are usually estimated using Monte Carlo hadronization models.
The renormalon inspired model of ref. \cite{Dokshitzer:1996qm}
provides an alternative method.
\subsection{Shape variables at highest energy}
\begin{figure}
\includegraphics[width=\figsize]{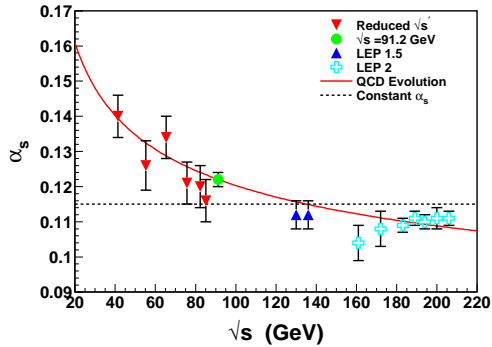}
\caption{L3 determinations of $\as$ from shape variables at various energies.}
\label{fig:L3-as-lep2}
\end{figure}

In ref. \cite{L32670:2001hep} we find an example of jet studies
at the highest LEP energies from the L3 experiment. They determine $\as$
from thrust, heavy jet mass, total jet broadening,
wide jet broadening and $C$ parameter, measured at different energies.
The results are shown in fig.~\ref{fig:L3-as-lep2}.
 Energies below the
$Z^0$ mass are reached via initial state photon radiation.
Their fitted value of the strong coupling
\begin{displaymath}
\as(M_Z)=0.1220\pm0.0011({\rm exp.}) \pm 0.0061({\rm th.})\;.
\end{displaymath}
is consistent with the world average.
The running of the QCD coupling is visible in the data.

\subsection{QCD color factors}
\begin{figure}
\includegraphics[width=\figsize]{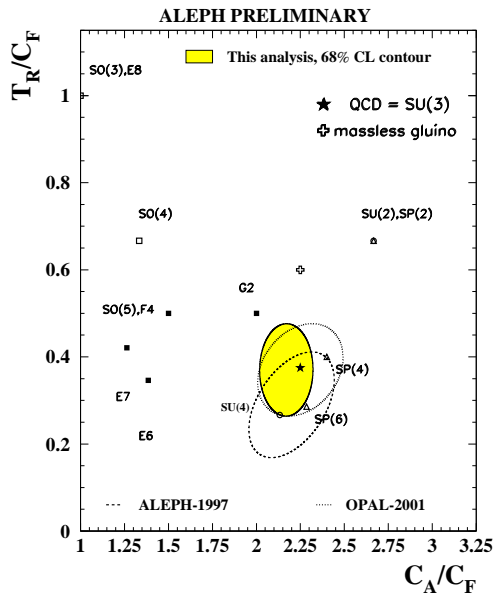}
\caption{Colour factors fits from ALEPH and OPAL.}
\label{fig:ALEPH-colfac1}
\end{figure}

The availability of NLO calculation for 4-jet rates has
made it possible to perform more reliable fits to the QCD
color factors from jet data.
Simultaneous fits to $\as$ and to the QCD colour factors have been
performed by the OPAL \cite{Abbiendi:2001qn} and the ALEPH
\cite{ALEPH2001-042} experiments. Both experiments use jet rates
to constrain the value of $\as$, and certain shape variables
that are sensitive to the angular distributions of the softest
jets.
\begin{figure}
\includegraphics[width=\figsize]{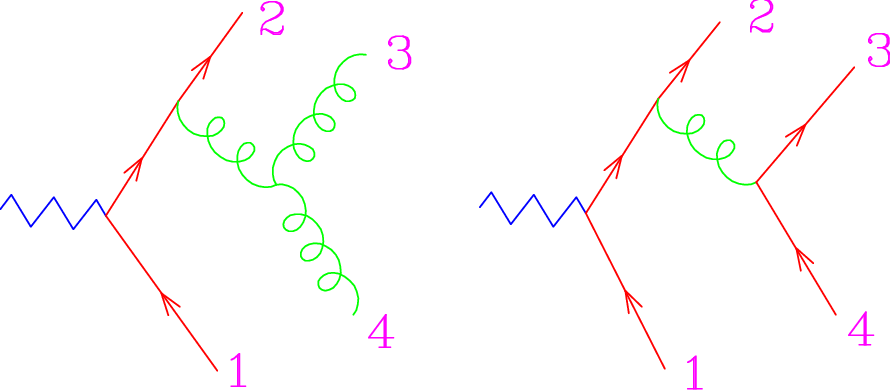}
\caption{Four jet production mechanisms.}
\label{fig:fourjets}
\end{figure}
As shown in fig.~\ref{fig:fourjets}, those are more likely to come
from the splitting of a radiated gluon (particle 3 and 4), and the
angular distribution of the gluon pair differs from that of the secondary
quark pair. The jet resolution parameter is
\begin{displaymath}
y_{ij}=2{\rm min}(E_i^2,E_j^2)(1-\cos\theta_{ij})/E_{\rm vis}^2
\end{displaymath}
and the variables considered are the differential 2-jet rate
$D_2(y_{23})=\frac{1}{\sigma_{\rm tot}}\frac{d\sigma}{dy_{23}}$ (only OPAL),
and the 4-jet rate $R_4(y_{\rm cut})$. Finally, they consider
the following variables constructed from 4-jet clusters
(defined with $y_{\rm cut}=0.008$, $E_1>E_2>E_3>E_4$)
\begin{eqnarray}
 \chi_{\rm BZ}&=&\angle[(\vec{p}_1\wedge  \vec{p}_2),
\,(\vec{p}_3\wedge  \vec{p}_4)] \nonumber \\
 \Theta_{\rm NR}&=&\angle[((\vec{p}_1 -  \vec{p}_2),
\,(\vec{p}_3 -  \vec{p}_4)] \nonumber \\
 \Phi_{\rm KSW}&=&\langle \angle[(\vec{p}_1\wedge  \vec{p}_4),
\,(\vec{p}_2\wedge  \vec{p}_3)]\rangle_{3\leftrightarrow 4} \nonumber \\
 \cos(\alpha_{34})&=&\cos(\angle[(\vec{p}_3,
  \vec{p}_4]) \nonumber
\end{eqnarray}
The color dependence enters the 2 and 4 jet rates in the form
 \begin{eqnarray}
D_2&=&\as \Cf \ldots \nonumber  \\
    &+&  \as^2 \Cf (\Cf\ldots + \Ca\ldots +\Tf \nf\ldots)
\nonumber \\
R_4&=&\as^2 \Cf (\Cf\ldots + \Ca\ldots +\Tf \nf\ldots) \nonumber \\
   &+& \as^3 \ldots \nonumber
\end{eqnarray}
Monte Carlo models are used to implement hadronization corrections.
The results are summarized in table~\ref{tab:ColourFit} and
fig.~\ref{fig:ALEPH-colfac1}.
\begin{table}
\begin{center}
\caption{Results for the simultaneous fit to $\as$, $\Ca$ and $\Cf$ from
OPAL and ALEPH.}
\label{tab:ColourFit}
{ \renewcommand\tabcolsep{1pt}
\begin{tabular}{|c|c|c|c|c|c|}
\hline\hline
     &       &        & stat.      & syst.      \\ \hline\hline
     & $\Ca$ & $3.02$ & $\pm 0.25$ & $\pm 0.49$ \\
 \cline{2-5} OPAL
     & $\Cf$ & $1.34$ & $\pm 0.13$ & $\pm 0.22$ \\
     \cline{2-5}
     & $\as(M_{\rm Z})$ & $0.120$ & $\pm 0.011$ & $\pm 0.020$ \\ \hline\hline
     & $\Ca$            & $2.93$  & $\pm 0.14$  & $\pm 0.49$  \\
 \cline{2-5} ALEPH
     & $\Cf$            & $1.35$  & $\pm 0.07$  & $\pm 0.22$ \\
     \cline{2-5}
     & $\as(M_{\rm Z})$ & $0.119$ & $\pm 0.006$ & $\pm 0.022$ \\
\hline\hline
\end{tabular}}
\end{center}
\end{table}
An older result from ALEPH, performed using four parton matrix elements
at leading order, is also shown. The new, NLO analysis show better
agreement with QCD expectations.
\subsection{Heavy Quark mass effects}
\begin{figure}
\includegraphics[width=\figsize]{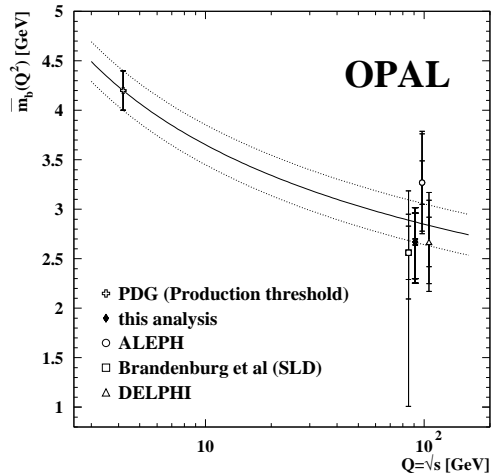}
\caption{Summary of $b$ mass determinations from $B_3$.}
\label{fig:OPAL-bmass}
\end{figure}

Comparison of light and heavy quark hadronic events exposes
the effects of the heavy quark mass. The analysis of this phenomenon,
is made possible by the availability of NLO computation of 3-jet cross sections
with massive quarks. One can fit the $b$ quark mass from the
ratio of 3-jet rates in $b$-quark jets and light quark jets
\begin{displaymath}
B_3=\frac{R_3^{b}}{R_3^{dusc}}
\end{displaymath}
Alternatively, one compares the $\as$ determination in light quark events and
$b$, $c$ quark events (tests of flavour independence of $\as$).
Results from $b$-mass fits to $B_3$ have been performed by various
experiments. A recent summary from the OPAL collaboration
ref.~\cite{Abbiendi:2001tw} is shown in fig.~\ref{fig:OPAL-bmass}.
Also DELPHI has performed a recent determination \cite{Bambade:2001hep},
reported here in table \ref{tab:DELPHImb}.
\begin{table}
\begin{center}
\caption{Results for $\bar{m_b}(M_Z)$ from DELPHI. The first column refers to the algorithm used to define jet rates: Durham or Cambridge.}
\label{tab:DELPHImb}
{
\renewcommand\tabcolsep{1pt}
\begin{tabular}{|c|c|c|c|c|c|}
\hline
&$\bar{m_b}(M_Z)$ & Stat. & Had. & Tag. & Th.  \\ \hline
\rule[-5pt]{0pt}{15pt}
Durh. & $2.67\,$GeV & $\pm0.25$ & $\pm0.34$ && $\pm 0.27$ \\ \hline
Camb. & $2.61\,$GeV & $\pm0.18$ & $\pm0.47$ &  $\pm 0.18$ & $\pm 0.12$ \\
\hline
\end{tabular}}
\end{center}
\end{table}
They argue that theoretical errors are much smaller
if jet rates are defined using the Cambridge algorithm.

Results on $B_3$ are often stated as evidence for the running of the $b$
quark mass. Whether this is in fact the case is debatable. It remains
however the fact that it is a measurement of the $b$ mass in high energy
processes, and that NLO corrections are essential to its success.
\section{Results from HERA}
The QCD program is prominent in the HERA experiments.
Structure function studies and small-x physics are
presented in separate reports \cite{Erdmann:LP2001,Carli:LP2001}.
Here I will discuss few recent results on jets studies.

Several recent publications deal with the photoproduction and DIS
production of jets at HERA. The DIS region is particularly interesting,
since it does not rely upon the knowledge of the photon parton densities.
Jets are reconstructed in the Breit frame, which is the frame where
the virtual photon is purely spacelike, and its momentum is antiparallel
to that of the incoming proton. In the parton model approximation,
the incoming proton, the virtual photon, and the outgoing parton
are all collinear in this frame.

\begin{figure}
\includegraphics[width=\figsize]{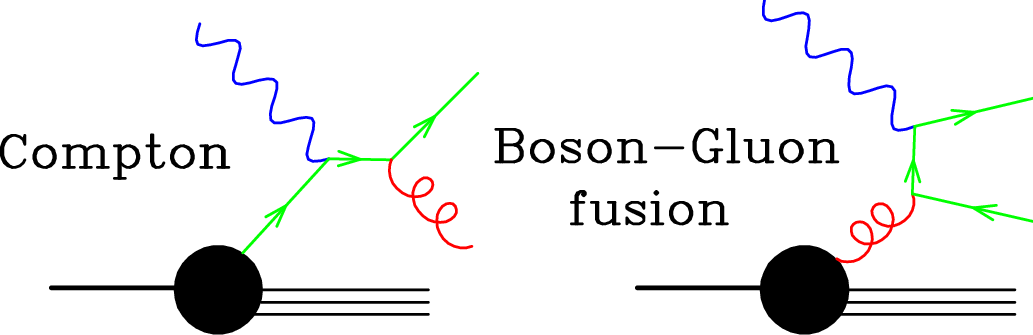}
\caption{Born level mechanisms for dijet production in DIS.}
\label{fig:HERAdijet}
\end{figure}

 Transverse momentum in the Breit frame is generated
with the mechanism depicted in fig.~\ref{fig:HERAdijet}.
Thus, transverse jet production in the Breit frame
is directly sensitive to $\as$. Jets can be reconstructed by a $k_T$
clustering algorithm applied in the Breit frame. These algorithms
are variants of those used in $e^+e^-$ physics,
that account for the presence of the beam remnant jet
\cite{Catani:1993hr,Ellis:1993tq,Wobisch:1998wt}.
Calculations at the next-to-leading order for dijet production
in DIS have been available for some time
\cite{Mirkes:1996ks,Graudenz:1997gv,Catani:1997vz}.

\begin{figure}
\includegraphics[width=\figsize]{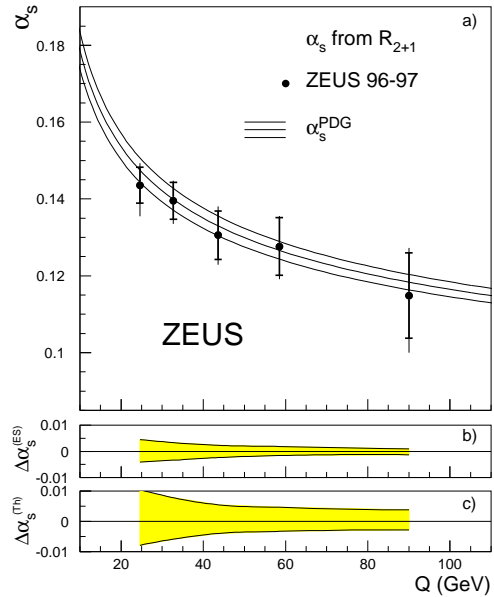}
\caption{Zeus determinations of $\as$ from dijet cross sections in DIS.}
\label{fig:Zeus-asdijet}
\end{figure}

\begin{figure}
\includegraphics[width=\figsize]{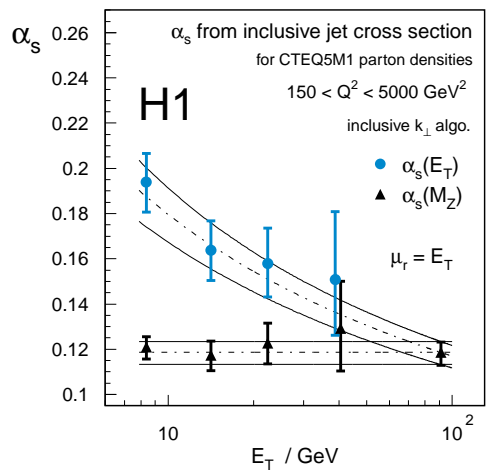}
\caption{H1 determinations of $\as$ from single inclusive jet production
in DIS.}
\label{fig:H1-asincjet}
\end{figure}

Zeus  \cite{Breitweg:2001rq,Chekanov:2001fw} and H1
\cite{Adloff:2000tq} have performed $\as$ studies using jets in DIS.
A summary of their analysis is shown in figs.~\ref{fig:Zeus-asdijet}
and \ref{fig:H1-asincjet} respectively.
The Zeus experiment uses the
2-jet fraction $R_{2+1}$, while H1 uses the inclusive jet cross section.
Thus, systematics and theoretical uncertainties are quite different
in the two methods. In spite of these differences, the two analysis
measure a consistent value of $\as$, as shown in table \ref{tab:ZeusH1as}.
\begin{table}
\label{tab:ZeusH1as}
\caption{Determination of $\as$ from Jets in DIS.}
{\renewcommand\tabcolsep{1pt}
\begin{tabular}{|c|c|c|c|c|c|}
\hline
& $\as(M_Z)$ & Stat. & Exp. & Th. & PDF \\ \hline
\rule[-5pt]{0pt}{15pt}
Zeus & $ 0.1166$ & $ \pm 0.0019$ & $ ^{+0.0024}_{-0.0033}$ &
\multicolumn{2}{c|}{$^{+0.0057}_{-0.0044}$} \\ \hline
\rule[-5pt]{0pt}{15pt}
H1 & $ 0.1186$& $ \pm 0.0007$& $ \pm 0.0030$&
$^{+0039}_{-0.0045}$ & $^{+0.0033}_{-0.0023}$ \\ \hline
\end{tabular}}
\end{table}
Furthermore, both experiment show some evidence for the running
of the strong coupling.

Besides being sensitive to $\as$, DIS jet cross
sections also depend upon the gluon density. In fact, some constraints
on the gluon density can be derived assuming a
fixed value of $\as$ from other determinations, or, alternatively,
$\as$ and the gluon densities can be fitted simultaneously
\cite{Adloff:2000tq}.
\subsection{3 jet study from H1}
\begin{figure}
\includegraphics[width=\figsize]{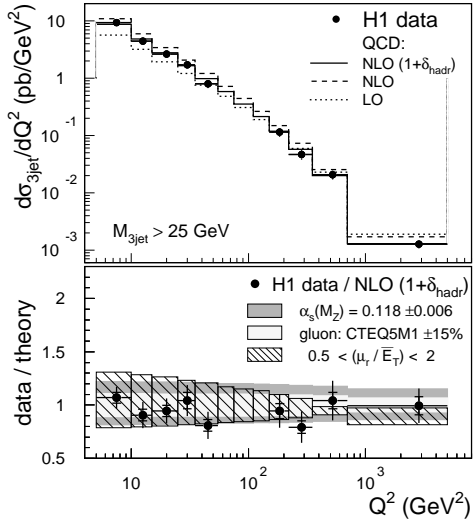}
\caption{Three-jet cross section in DIS, as a function of $Q^2$.}
\label{fig:H1-3jetdist}
\end{figure}

\begin{figure}
\includegraphics[width=\figsize]{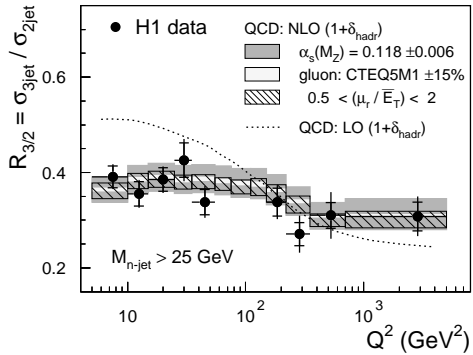}
\caption{Three jet to two-jet ratio compared to theoretical
predictions.}
\label{fig:H1-3-2-rat}
\end{figure}

Very recently, NLO calculations for 3-jet production have
become available \cite{Nagy:2001xb}. This has allowed the H1
collaboration to perform for the first time a NLO study of 3-jet
production at HERA \cite{Adloff:2001kg}.
Jets are defined with a $k_T$ cluster algorithm.
A minimum jet transverse energy of 5~GeV (in the Breit frame) is required,
and the three (two) jet sample consists of all events with three (two)
jets with invariant mass above 25~GeV. Further pseudorapidity cuts
are imposed in the laboratory frame.
In fig.~\ref{fig:H1-3jetdist} the three jet cross
section is shown with a detailed comparison of theoretical prediction
versus data, showing good agreement.
In fig.~\ref{fig:H1-3-2-rat}
the ratio of the three to two jet cross section is displayed.
The presence of NLO effects is clearly visible in the figure.
Studies of this kind, performed with higher statistics, have a good
potential for $\as$ determination, since the parton density uncertainties
partly compensate in the ratio.
\section{Hadron collider physics}
Inclusive jet production at colliders has been intensively studied in the
past decade. NLO results for the cross section have been available for a long
time \cite{Ellis:1990ek,Aversa:1990uv,Giele:1994gf}, and much work has been
spent in refining jet definitions appropriate for the collider environment.
\subsection{Single inclusive jet cross section}
\begin{figure}
\includegraphics[width=\figsize]{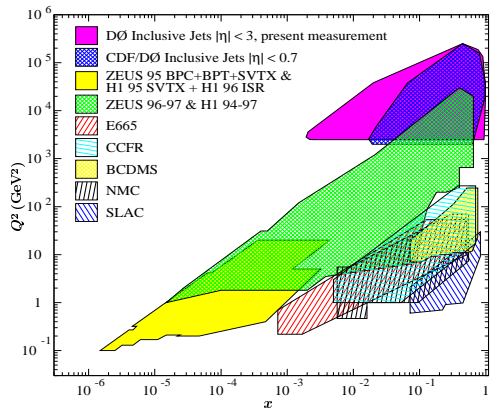}
\caption{The reach of the D0 inclusive jet analysis in the
$Q^2,x$ plane for the parton densities.}
\label{fig:D0-pdf-reach}
\end{figure}

\begin{figure}
\includegraphics[width=\figsize]{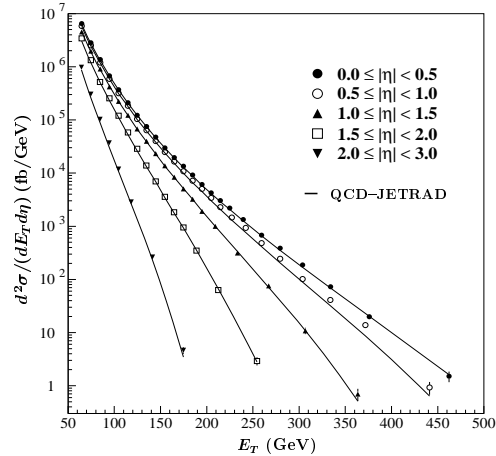}
\caption{Inclusive jet cross section
as a function of $E_T$, in various rapidity bins, versus
theoretical predictions.}
\label{fig:D0-incjet-rap}
\end{figure}

In ref.~\cite{Abbott:2000ew}, a recent measurement of inclusive jet
cross section in a wide rapidity range is reported.
By exploring the high rapidity region, one extends toward smaller
values of $x$ the region
in the $Q^2,x$ plane where parton densities are probed, as shown
in fig.~\ref{fig:D0-pdf-reach}.
Jets are defined with the usual $\eta\phi$
iterative cone algorithm, with a radius $R=0.7$.
The D0 results, together with a NLO QCD predictions, are shown in
fig.~\ref{fig:D0-incjet-rap}, showing a remarkable agreement.
A more detailed comparison is shown in fig.~\ref{fig:D0-jet-th},
where the ratio $(\mbox{data}-\mbox{theory})/\mbox{theory}$ is plotted.
Theoretical results are obtained with the program JETRAD \cite{Giele:1994gf},
using the CTEQ4 \cite{Lai:1997mg} (left figure) and MRST \cite{Martin:1998sq}
(right figure) structure functions.
\begin{figure*}
\begin{center}
\includegraphics[width=\figsize]{D0-jet-th1.seps}
\includegraphics[width=\figsize]{D0-jet-th2.seps}
\caption{
Comparison of experimental measurements versus theoretical predictions:
CTEQ4HJ ({\boldmath$\bullet$}) and CTEQ4M ({\boldmath$\circ$})
(left figure);
MRSTg$\uparrow$ ({\boldmath$\bullet$}) and MRST ({\boldmath$\circ$})
(right figure).}
\label{fig:D0-jet-th}
\end{center}
\end{figure*}
The shaded band corresponds to one standard deviation on the systematic error.
One expects a comparable band for the theoretical error. The data is therefore
in good agreement with theoretical predictions, showing a preference
for the CTEQ4 sets.
\subsection{Dijet cross sections}
\begin{figure}
\includegraphics[width=\figsize]{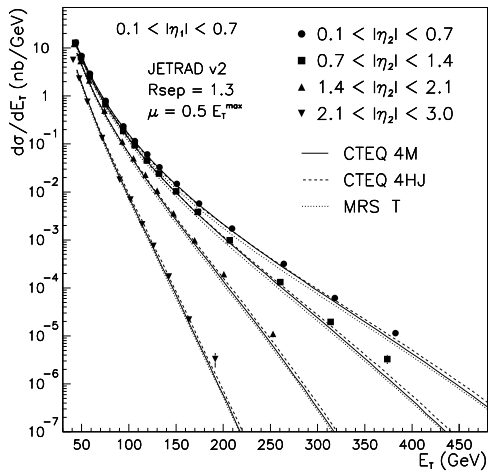}
\caption{Dijet cross sections from CDF; $E_T$ distribution of one
central jet, for the recoiling jet in different rapidity bins.}
\label{fig:cdf-dijetfig}
\end{figure}

\begin{figure}
\includegraphics[width=\figsize]{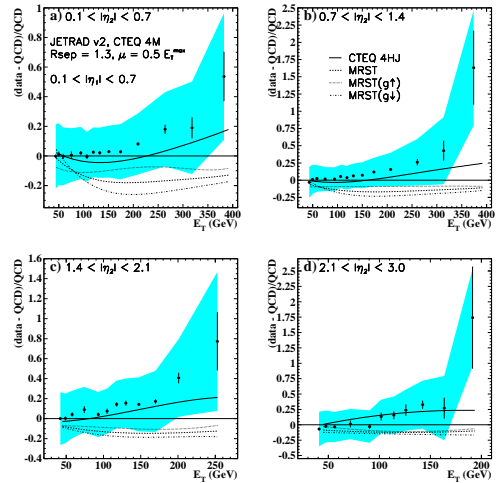}
\caption{Comparison if the dijet cross section to theoretical predictions.
The error bars represent the statistical errors, while the shaded
band represents the correlated systematic error.}
\label{fig:cdf-dijet-th}
\end{figure}

CDF has performed a study of dijet production \cite{Affolder:2000ew}.
They look at the $E_T$ of one central jet ($0.1<\eta_1<0.7$), while
the second jet lies in several different
pseudorapidity intervals. In this way, the sensitivity to the
parton densities at large $x$ is enhanced.
Qualitatively the theory gives a good description of data, as
can be seen from fig~\ref{fig:cdf-dijetfig}.
A closer look reveals problems at the quantitative level.
Looking at the $(\mbox{data}-\mbox{theory})/\mbox{theory}$ ratio
in fig.~\ref{fig:cdf-dijet-th},
one sees that no parton density functions set fits the data
satisfactorily, especially in the high $E_T$ region.

We recall that jet studies at the Tevatron is at the frontier
of our knowledge on the parton density functions. In fact,
the single inclusive jet cross
section \cite{Affolder:2001fa} was found initially to be higher
than QCD predictions. Further studies have shown that the
excess over perturbative predictions is within
the current flexibility in our parametrization of the parton density.
However, more detailed studies may reveal further problems.
\section{Power Corrections}
\begin{figure}
\includegraphics[width=\figsize]{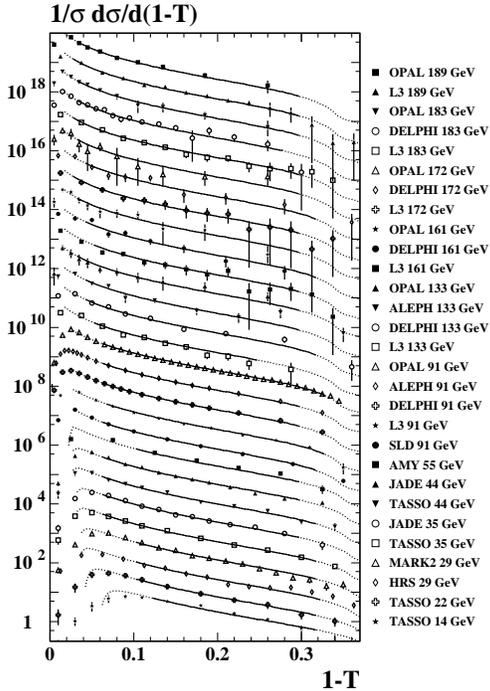}
\caption{Fit to thrust distributions from experiments at different energies,
using perturbation theory plus power corrections.}
\label{fig:OPAL-124-thrust}
\end{figure}

\begin{figure*}
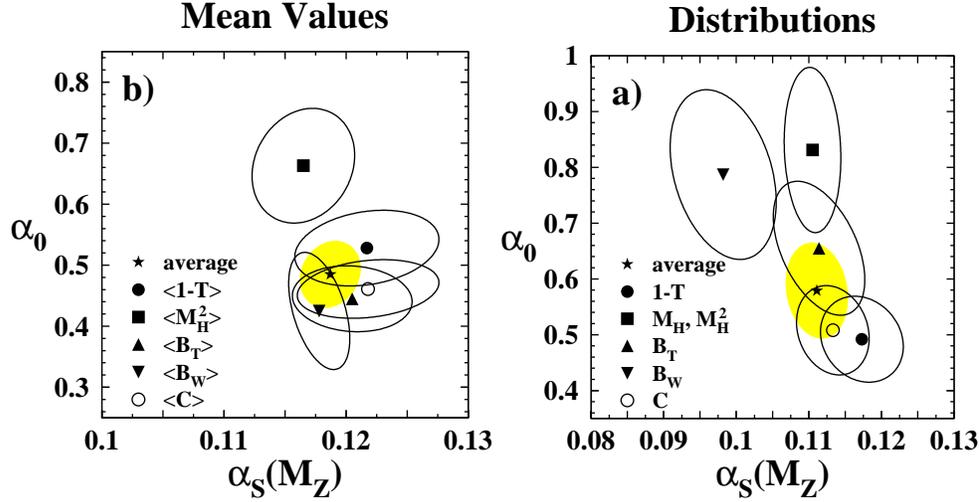

\includegraphics[width=\figsize]{OPAL-124-a0fitsav.seps}
\includegraphics[width=\figsize]{OPAL-124-a0fitsdist.seps}
\caption{Simultaneous fits to $\as$ and $\alpha_0$ using mean values
of shape variables (left) and distributions (right).}
\label{fig:OPAL-124-a0fit}
\end{figure*}

In most shape variables studies, hadronization corrections
are removed from the data using a Monte Carlo model. Since Monte Carlo
parameters are tuned to fit the data, this creates a complex interrelation
between what one predicts and what one measures. It would be desirable
to have power correction models which are simpler (and have better theoretical
motivation) than those used in Monte Carlo programs. Some activity
in this direction has started, especially following the work of
ref.~\cite{Dokshitzer:1996qm}.
In ref.~\cite{MovillaFernandez:2001ed}, several shape variables have been examined in the energy range of $\sqrt{S}=14$ to $189$ GeV. QCD NLO prediction,
together with power corrections are used to fit the data.
In the model of ref.~\cite{Dokshitzer:1996qm}, the leading power
corrections to shape variables are controlled by a single
non-perturbative parameter $\alpha_0$. Thus, the parameters of the fits
are $\as$ and $\alpha_0$.
In fig.~\ref{fig:OPAL-124-thrust}, the results for the fit to the thrust
distribution for the different data sets considered is displayed.
The agreement, in the very large energy range considered, is quite
remarkable. A summary of the results of this analysis is displayed
in figs.~\ref{fig:OPAL-124-a0fit} and table \ref{tab:Movilla}.
\begin{table}
\label{tab:Movilla}
\caption{Results of the fits to $\as(M_Z)$ and $\alpha_0(2\,{\rm GeV})$
from ref.~{\protect \cite{MovillaFernandez:2001ed}}.}
{\renewcommand\tabcolsep{1pt}
\begin{tabular}{|c|c|c|c|c|c|c|c|c|c|}
\hline
 & & & fit & syst. & Th.\\ \hline\hline
\rule[-5pt]{0pt}{15pt} means
 & $\as$ & $0.1187 $ & $\pm0.0014$ & $\pm0.0001$ & $^{+0.0028}_{-0.0015}$
 \\ \cline{2-6}
& $\alpha_0$ & $0.485  $ & $\pm 0.013$ & $\pm0.001$  & $^{+0.065}_{-0.043}$
 \\ \hline\hline
\rule[-5pt]{0pt}{15pt} distr.
 & $\as$ & $0.1111 $ & $\pm0.0004$ & $\pm0.0020$ & $^{+0.0044}_{-0.0031}$
 \\ \cline{2-6}
& $\alpha_0$ & $0.579  $ & $\pm 0.005$ & $\pm0.011$  & $^{+0.099}_{-0.071}$
 \\ \hline\hline
\rule[-5pt]{0pt}{15pt} Comb.
 & $\as$ & $0.1171 $ & \multicolumn{3}{c|}{$^{+0.0032}_{-0.0020}$}
 \\ \cline{2-6}
\rule[-5pt]{0pt}{15pt}
& $\alpha_0$ & $0.513  $ & \multicolumn{3}{c|}{$^{+0.066}_{-0.045}$}
 \\ \hline\hline
\end{tabular}}
\end{table}
We observe that the final value is well in agreement with 
other determinations \cite{Bethke:2000ai}. On the other hand, the value
determined from distributions is considerably lower than the value
obtained with standard methods (i.e. hadronization corrections with
Monte Carlo models). Furthermore, for some shape variables the consistency
of the determination is quite poor.
Studies of this kind have also been performed by the H1 and Zeus experiments
\cite{Poeschl:Moriond2001}.

Current knowledge on power corrections is somewhat limited, although there
are a few things that are better established.
For example, several theoretical arguments, based upon the Operator
Product Expansion and the structure of Infrared Renormalons, give us
confidence that in reactions similar to $R_{e^+e^-}$, power
corrections behave as $1/Q^4$. This fact is also supported by the
experimental studies on the $\tau$ spectral functions.
In DIS, one expects $1/Q^2$ corrections for the twist expansion,
and again one observes scaling phenomena at relatively low $Q^2$.
From several points of view, we are convinced that power corrections
in jets behave as $1/Q$. This means that may still be important
even at LEP energies, and some modeling of these effects may be needed
in order to describe the data satisfactorily.

From a theoretical point of view, power suppressed effects is an extremely
difficult topics.
Even in the simplest case of $R_{e^+e^-}$, where
the tools of the Operator Product Expansion is applicable, it has been
argued that the correspondence of a $1/Q^4$ correction in the OPE and
in the physical quantity may be broken \cite{Shifman:2000jv} in the process
of continuation from the euclidean to the physical region.
Furthermore, several argument show that power corrections are connected
to the divergence of the perturbative expansion, and thus non-perturbative
corrections are unambiguously defined only when a procedure for the
summation of the (asymptotic) perturbative expansion has been defined.
In other words, the there is no sound justification to adding non-perturbative
corrections to a truncated perturbative expansion. All these problems are
even worse in the case of jets, where not even a
formulation of the problem in terms of the Operator Product Expansion
is available.

Current models basically assume that power corrections arise
from the integration of the strong coupling constant over
the low momentum region. Physical quantities of order $\as$ are
written in terms of the running coupling as:
 \begin{displaymath}
F(Q^2)=\int dk^2 {\cal F}(Q^2, k^2) \as(k^2)\;.
\end{displaymath}
where ${\cal F}(Q^2, k^2)$ is the physical quantity computed with
a gluon of mass $k^2$.
The low energy behaviour of $\as$ determines power corrections.
The power correction law is determined by the low $k^2$ behaviour of
${\cal F}(Q^2, k^2)$. In particular
\begin{displaymath}
F(Q^2)\propto \frac{1}{k} \quad \to \quad \mbox{$1/Q$ correction}
\end{displaymath}
This is the starting point of the model of ref.~\cite{Dokshitzer:1996qm}.
It implies that power corrections are modeled in different shape variables
by the same unknown parameter, which is a weighted integral
of the strong coupling constant over the small momentum region.
It is unclear whether this factorization hypothesis can survive
higher order corrections
\cite{Nason:1995hd,Dokshitzer:1998iz,Dokshitzer:1998pt}.
The model has a single non-perturbative parameter, and the
data, to some extent, supports its validity. It has also been
used to perform a fit to the QCD colour factors \cite{Kluth:2000km},
since it allows for a colour dependent parametrization of the power
suppressed corrections.

There is an interplay between power corrections and unknown
higher order effects. On one hand, one fits a fixed order formula,
which is valid up to inverse powers of the logarithm of the scale,
plus a term which behaves like an inverse power of the scale, and thus is
formally smaller than the unknown higher order terms. Prescriptions
that try to improve the perturbative expansion by some guess of its
higher order behaviour (similar to the principle of minimal sensitivity)
sometimes can mimic the effect of power suppressed terms. For example,
in ref.~\cite{DELPHI166:LP2001}, the average values of several shape variables
is fitted in this way, without any need for the introduction of power
suppressed effects. The inclusion of next-to-next-to-leading order effects
in the computation of shape variables may help to clarify this issue
in the future.
\subsection{Status of the NNLO calculations}
Although today's typical QCD calculations include terms up to the
NLO level, there are a few classes of problems where higher order
terms have been computed. Among the most important results:
  the total cross section for $e^+e^- \to {\rm hadrons}$, computed to
  order $\as^3$ \cite{Gorishnii:1988bc,Gorishnii:1991vf};
  the QCD $\beta$ function, computed at the 4-th loop level
  (${\cal O}(\as^5)$) \cite{vanRitbergen:1997va};
  up to $N=12$ singlet, $N=14$ non-singlet crossing even,
  and $N=13$ crossing odd moments of the
  ${\cal O}(\as^3)$ splitting functions, together with
  the ${\cal O}(\as^3)$ coefficient functions for DIS \cite{Larin:1997wd}
  \cite{Retey:2000nq} \cite{vanNeerven:1991nn}.
In all these cases, the problem can be reduced to the computation
of a massless propagator type graph, which can be computed with the
techniques developed in refs. \cite{Gorishnii:1989gt,Larin:1991fz}.
These results have important consequences on $\as$ determination
from $Z$ and $\tau$ decays, and in DIS processes.

The only collider process that has been computed to NNLO is the
Drell-Yan pair production cross section
\cite{Hamberg:1991np}. This process is particularly simple at the Born
and NLO level.
NNLO calculations in typical
collider processes (which already have a certain complexity at the NLO
level) are extremely challenging. Several research groups are
working on particular aspects of these calculations.
The focus is on jet production in hadronic collision, that is
to say, on the parton-parton scattering process
(jet production in $e^+e^-$ annihilation, which would be
particularly useful for LEP physics, is next in complexity
\cite{Weinzierl:2001hep}, since it
depends upon one more kinematic invariant).
In order to compute the NNLO jet production cross section one needs:
{\blue \begin{itemize}
\item
the square of the {\red $2\to 4$} tree amplitude;
\item
the interference of the {\red $2\to 3$} tree level and one loop amplitude;
\item
the square of the {\red $2\to 2$} one loop amplitude;
\item
the interference of the {\red $2\to 2$} two loop and tree level amplitude;
\item
for a consistent phenomenological treatment, the ${\cal O}(\as^3)$
Altarelli-Parisi splitting functions are also needed.
\end{itemize}}
For the last item, an
approximate expression, based upon constraints coming from the
large and small $x$ behaviour, and from the moments of the splitting functions
known at the NNLO level, has been obtained~\cite{vanNeerven:2001pe}.
A calculation of the full NNLO splitting function is under way
\cite{Moch:2001fr}.

Techniques for the computation of the tree level amplitudes for the $2\to 4$
process have been available for a long time. The problem is the collinear
and soft limits of such amplitudes, that must be regularized in order
to implement the cancellation and factorization of soft and collinear
divergences. In dimensional regularization, these singularities will
appear as poles in $1/\epsilon$ up to the fourth power. Therefore,
the structure of these singularities must be understood with the
required accuracy in $\epsilon$, in order to get a correct result
after cancellation. All these results are available now:
{\blue
\begin{itemize}
\item
Double soft limit \cite{Berends:1989zn}:
when two final state particles become soft. Can yield singularities
up to {\red $1/\epsilon^4$} after final state integration of the soft
particles.
\item
Double collinear and soft-collinear
\cite{Campbell:1998hg,Catani:1999ss,DelDuca:1999ha}:
a subset of 3 final state partons become collinear, or 2 become
collinear and one is soft (up to $1/\epsilon^3$ singularities).
\end{itemize}}
The $2\to 3$ amplitudes at one loop level are also known
\cite{Bern:1993mq,Kunszt:1994tq}.
Again, the collinear and soft limit of this amplitude is needed in order
to meet the accuracy required by NNLO calculations:
{\blue
\begin{itemize}
\item
collinear limit of one loop amplitudes
\cite{Bern:1994zx,Kosower:1999xi,Kosower:1999rx};
\item
soft limit of one loop amplitudes
\cite{Bern:1998sc,Bern:1999ry,Catani:2000pi}.
\end{itemize}}
The two loop $2\to 2$ contribution has been recently computed
\cite{Anastasiou:2000kg,Anastasiou:2000ue,Glover:2001af}.
Recursion relations among Feynman integrals are used to reduce the
problem to the computation of a small number of master integrals.
The hardest of those (double box, planar and crossed) have been solved
only recently \cite{Smirnov:1999gc,Tausk:1999vh}
The structure of the $1/\epsilon$ singularities of these amplitude
can be checked against a general factorization formula
\cite{Catani:1998bh}.

All ingredient needed for a full NLO computation of jet cross
sections in hadronic collisions are in place.
The implementation of the various terms into a useful result
still appears as a formidable task.
However, in view of the enormous progress achieved in the last few years,
it appears now that a result will become available in useful time.
\section{Puzzle in $b$ production}
It has been known for many years that the $b$ production cross section
measured at hadron colliders is above QCD prediction by roughly a factor
of 2 (see, for example \cite{Gutierrez:2001hep}).
This discrepancy has never been considered too worrisome, since
NLO corrections to $b$ production are of the order of 100\%, and thus
NNLO effects could be of the same order. The effect of Sudakov resummation
\cite{Bonciani:1998vc},
high transverse momentum resummation \cite{Cacciari:1998it},
and small-$x$ resummation \cite{Collins:1991ty}
give individually small contributions, of the order of 10 to 30\%.
However, they are all positive, and thus
it is not unlikely that this discrepancy may be explicable by a cocktail of
several different effects.

More serious discrepancies have been observed by the H1 and Zeus
experiments in the past, in the context of heavy flavour photoproduction.
They gave an indication of cross sections larger than NLO QCD prediction
by more than a factor of 2. In this case, the theoretical prediction
are more solid, since the process of photoproduction is partly electromagnetic,
and it has therefore smaller strong corrections.

Very recently, an excess in the $b$ production cross section has also
been observed in $\gamma\gamma$ collisions.

\begin{figure}
\includegraphics[width=\figsize]{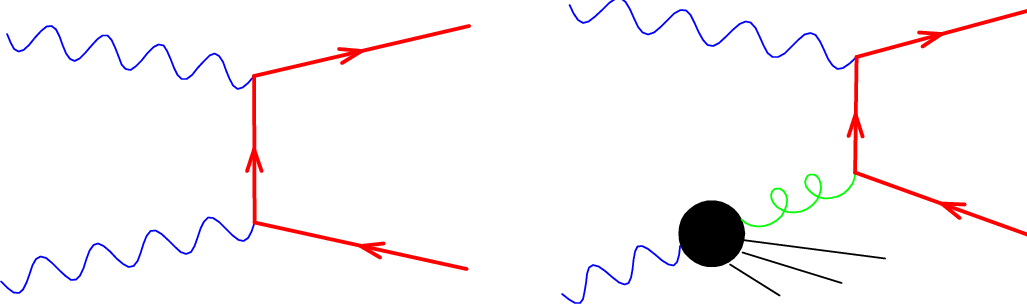}
\caption{Heavy flavour production mechanism in $\gamma\gamma$
collisions: direct photon fusion process, and single resolved process.}
\label{fig:gg-b.eps}
\end{figure}

The heavy flavour production mechanism in $\gamma\gamma$
collisions is depicted in fig.~\ref{fig:gg-b.eps}.
At LEP energies, the direct and single resolved processes give contributions
of the same order, while the double resolved process is negligible
\cite{Drees:1993eh}.
The L3 experiment has performed a measurement of the
$e^+e^-\to e^+e^- b\bar{b} X$ cross section \cite{Acciarri:2000kd}
using the electrons
or muons from the semileptonic $b$ decays. The $b$ cross section
is extracted by fitting the transverse momentum of the lepton
relative to the closest hadronic jet. They get
\begin{displaymath}
\sigma(e^+e^-\to e^+e^- b\bar{b} X)\mu = 14.9\pm 2.8 \pm 2.6 {\rm pb}\;,
\end{displaymath}
\begin{displaymath}
\sigma(e^+e^-\to e^+e^- b\bar{b} X)e = 10.9\pm 2.9 \pm 2.0 {\rm pb}\;.
\end{displaymath}
The OPAL experiment (using only $\mu$'s) gets comparable
results \cite{OPALNote:2000sc}.

\begin{figure}
\includegraphics[width=\figsize]{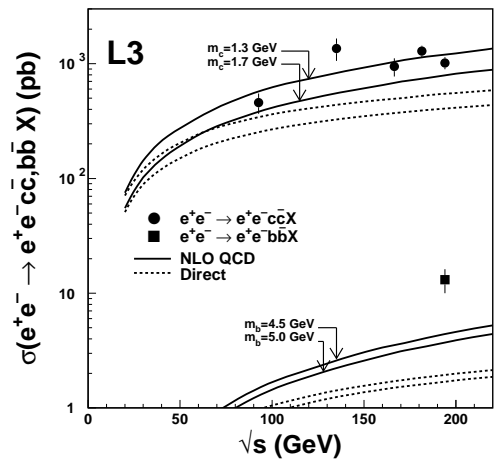}
\caption{L3 results for heavy flavour production in $\gamma$--$\gamma$
collisions.}
\label{fig:L3-gg-b-fig}
\end{figure}

\begin{figure}
\includegraphics[width=\figsize]{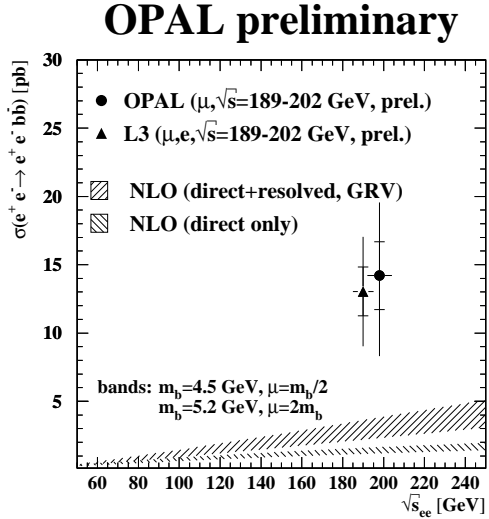}
\caption{OPAL result for $b$ production in $\gamma$--$\gamma$
collisions.}
\label{fig:OPAL-gg-b-fig}
\end{figure}

In fig.~\ref{fig:L3-gg-b-fig} the L3 results are reported. One can clearly
see that the $c$ cross section is compatible with QCD predictions
(see also \cite{Acciarri:2000sc}),
while the $b$ cross section is in excess.
The OPAL result is shown in fig.~\ref{fig:OPAL-gg-b-fig},
compared to a QCD calculation including some
more realistic estimates of the theoretical errors. Again, the discrepancy is
quite evident.

\begin{figure}
\includegraphics[width=\figsize]{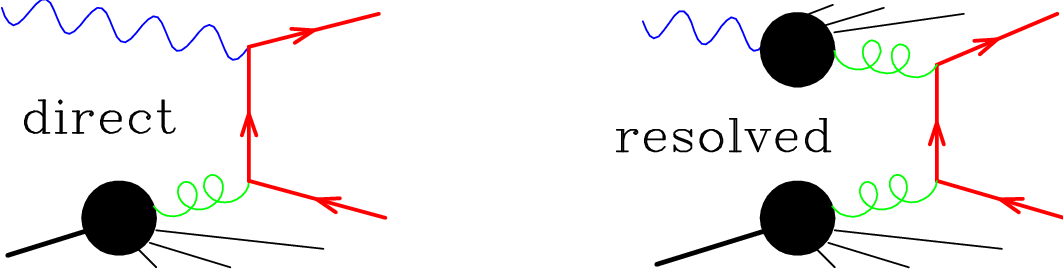}
\caption{QCD mechanism for the photoproduction of heavy flavours.}
\label{fig:pg-b}
\end{figure}

\begin{figure}
\includegraphics[width=\figsize]{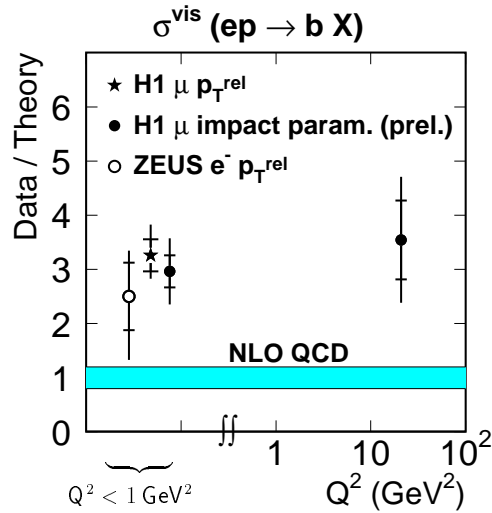}
\caption{Summary of experimental results for $b$ photoproduction, from
ref.~{\protect \cite{H1484:LP2001}}.}
\label{fig:H1prelim-b-fig}
\end{figure}

New results have also become available for $b$ photoproduction.
The mechanism for photoproduction of heavy flavours is depicted in
fig.~\ref{fig:pg-b}.
NLO calculations are available both for the direct and resolved components
\cite{Nason:1988xz,Ellis:1989sb,Beenakker:1991ma,Smith:1992pw}.
The resolved component of the cross section is moderate,
especially if one focus upon the central production region
\cite{Frixione:1997ma}. The experimental situation is summarized
in fig.~\ref{fig:H1prelim-b-fig} from ref.~\cite{H1484:LP2001}.
Previous results where obtained from fits to the relative transverse
momentum of muons \cite{Adloff:1999nr} and electrons \cite{Breitweg:2000nz}.
The new H1 results are obtained also from the impact parameter of
secondary vertices in semileptonic decays. Furthermore,
H1 measures also $b$ production in DIS with the same method.
The New {\red H1} results confirm previous findings, of an excess
in the $b$ production cross section by more than a factor of 2.

In view of the fact that theoretical uncertainties seem to be small
in the photoproduction process, it is unlikely that these discrepancies
may be explicable by unknown higher order terms, or non-perturbative
effects.
\section{Conclusions}
Progress in the field of high energy QCD is constantly
being made. At present, QCD has been tested in several different
area, with considerable success. Most studies involve QCD calculation
at the NLO level. A considerable effort is currently being made
to carry out next-to-next-to-leading order calculations.

From a phenomenological viewpoint, QCD describes quite well
strong interactions phenomena in high energy collisions. There are
a few area of poor agrement, that can be, in general, attributed
to large unknown higher order effects, or to non-perturbative physics.
One noticeable exception is $b$ production in $\gamma\gamma$ and $ep$
collisions, which, as of now, exhibit a disagreement with theoretical
predictions that seems to be too large to be explicable in terms 
of higher order effects or non-perturbative physics.

\section{Questions}
\begin{itemize}
\item[Q.] Mike Albrow, Fermilab:\newline
One should mention that there has also been an excess of high $p_T$
b-jets at the Tevatron with respect to NLO QCD.
\item[A.]
Yes, this discrepancy has been around for a long time. However,
in the case of hadronic collisions, QCD radiative corrections are
very large, and it is not unlikely that higher order effects are even larger.
\item[Q.] Tancredi Carli, Desy:\newline
Concerning the $b$ excess in $\gamma p$ and DIS,
it will be shown in the small-$x$ talk, that there are ideas emerging
from analysis of unintegrated gluon densities at HERA which are able
to get closer to the data and in addition get the $b$ cross section
in proton-antiproton collisions correctly.
\item[A.:] 
The question is whether the approaches you mention include NLO terms
correctly. I believe they do not.
\item[Q.] Guido Altarelli, CERN:\newline
A very naive explanation of the $b$ excess would be $b$ versus $c$
misidentification. Thus the dependence of the effect on the parameters
of the identification procedure would be interesting.
\item[A.]
The method that has been mostly used is based upon the $p_T$ of the
lepton relative to the closest jet, which has a different shape
for $b$, $c$ and other sources of leptons.
The recent addition of the results obtained using also the impact parameter,
by the H1 collaboration, confirms the effect. Presumably,
the HERA future runs will definitely clarify the issue.
\item[Q.] G. Iacobucci, INFN Bologna:\newline
The $p_T$ spectrum of charm is much softer than for beauty, and agrees with
NLO calculations for charm.
\item[Q.] John Ellis, CERN:\newline
Is there any information about the distribution of the excess of $b\bar{b}$
events? Are they mainly 2-jet or 3-jet configurations?
\item[A.]
The analysis published so far have examined only total rates.
\end{itemize}
\providecommand{\href}[2]{#2}\begingroup\raggedright\endgroup

\end{document}